%% file: main.tex
\documentclass[10pt,conference]{IEEEtran}

\input{setup}

\IEEEoverridecommandlockouts
\usepackage{cite}
\usepackage{amsmath,amssymb,amsfonts}
\usepackage{algorithmic}
\usepackage{graphicx}
\usepackage{textcomp}
\usepackage{soul}
\usepackage{xcolor}
\usepackage[]{footmisc}
\usepackage{mwe}
\usepackage{tabularx}
\def\BibTeX{{\rm B\kern-.05em{\sc i\kern-.025em b}\kern-.08em
    T\kern-.1667em\lower.7ex\hbox{E}\kern-.125emX}}

\usepackage{booktabs}

\usepackage{hyperref}

\newboolean{showcomments}
\setboolean{showcomments}{true} 
\ifthenelse{\boolean{showcomments}}
    {
        \newcommand{\kimberly}[1]{\textcolor{blue}{{\it [KL: #1]}}}
        \newcommand{\shurui}[1]{\textcolor{red}{{\it [SZ: #1]}}}
        \newcommand{\safwat}[1]{\textcolor{green}{\it SH: #1}}
    }
    {
        \newcommand{\kimberly}[1]{}
        \newcommand{\shurui}[1]{}
        \newcommand{\safwat}[1]{}
    }

\begin{document}

\title{LLMs in Mobile Apps: Practices, Challenges, and Opportunities
}

\author{\IEEEauthorblockN{Kimberly Hau}
\IEEEauthorblockA{\textit{University of Toronto} \\
kimberly.hau@mail.utoronto.ca}
\and
\IEEEauthorblockN{Safwat Hassan}
\IEEEauthorblockA{\textit{University of Toronto} \\
safwat.hassan@utoronto.ca}
\and
\IEEEauthorblockN{Shurui Zhou}
\IEEEauthorblockA{\textit{University of Toronto} \\
shuruiz@ece.utoronto.ca}
}
\maketitle
\begin{abstract}
The integration of AI techniques has become increasingly popular in software development, enhancing performance, usability, and the availability of intelligent features. With the rise of large language models (LLMs) and generative AI, developers now have access to a wealth of high-quality open-source models and APIs from closed-source providers, enabling easier experimentation and integration of LLMs into various systems. This has also opened new possibilities in mobile application (app) development, allowing for more personalized and intelligent apps. However, integrating LLM into mobile apps might present unique challenges for developers, particularly regarding mobile device constraints, API management, and code infrastructure. In this project, we constructed a comprehensive dataset of 149 LLM-enabled Android apps and conducted an exploratory analysis to understand how LLMs are deployed and used within mobile apps. This analysis highlights key characteristics of the dataset, prevalent integration strategies, and common challenges developers face. Our findings provide valuable insights for future research and tooling development aimed at enhancing LLM-enabled mobile apps.

\end{abstract}


\section{Introduction}
\looseness = -1
Recent years have not only witnessed significant advancements in natural language processing (NLP) tasks, but also launched large language models (LLMs) into recognition beyond its niche in NLP. Since then, the area has been moving fast~\cite{minaee2024large}, 
with LLMs showing potential toward becoming the basic building block for general-purpose artificial intelligence (AI) agents, and generative AI. Characterised as transformer-based language models with over a billion parameters and pre-trained on large Web-scale text corpus, it has been found that scaling up both the size of training data and parameter count of LLMs increases the capability of the models to approximate human-level performance in various tasks~\cite{naveed2023comprehensive}.

\looseness=-1
As such, LLMs are now frequently integrated into various types of software to enhance functionality across a wide array of applications \cite{minaee2024large,zhao2023survey, meyer2023chatgpt, fan2023large, yang2024if}. 
Similarly, mobile applications (apps)  are increasingly incorporating LLM components, bringing advanced AI capabilities directly to users' devices. However, deploying LLMs to mobile devices presents significant challenges. 
Prior studies have explored the technical aspects of the integration of AI models into products, 
and observed that the resource-constrained restrictions of mobile devices make LLMs especially difficult to deploy locally to mobile apps~\cite{carreira2023revolutionizing}. 
As studied by Alizadeh et al., the unprecedented capabilities of LLMs come with substantial computational and memory requirements for inference~\cite{alizadeh2023llm}. Also,  Çöplü et al. investigated the performance of LLMs on various smartphones and concluded that 
further advancements in power management and system integration are required for achieving sustained performance on even top-of-the-line mobile devices~\cite{ccoplu2023performance}. 
This is corroborated by Murthy et al., who benchmarked the performance of LLMs on on-device use cases and concluded that current LLMs require significant resources in terms of CPU and RAM usage when deployed on mobile devices~\cite{murthy2024mobileaibench}.
In addition to deploying LLMs on-device, another option 
for LLM deployment is vendor-provided APIs of commercial LLMs, such as Google’s Gemini ~\cite{gemini} and OpenAI’s GPT ~\cite{chatgpt}, which allows users the ability to run inference on closed-source LLMs without requiring memory and computational costs of storing and running the model locally. However, as observed by Minaee et al., this comes with the costs of maintaining subscriptions to the LLM vendor, latency lags, and security concerns with transmitting information~\cite{minaee2024large}.

Overall, there remains a need for an examination of the methods commonly employed by practitioners to integrate LLMs into Android apps, along with an analysis of the distinct challenges associated with each approach.
Therefore, in this study, we aim to provide insight into the practices, opportunities, and challenges 
that surround the current deployment of LLMs to Android apps. This includes investigating characteristics of LLM-enabled apps, integration strategies for merging LLMs with traditional code, handling the challenges of maintenance and updating apps in time with LLM updates, and other common concerns with deploying LLMs to mobile apps. For this project, we define \textbf{LLM-enabled mobile apps} as mobile apps that contain one or more LLM components and use such components to provide functionalities.
We compiled a detailed dataset comprising 149 Android apps available on GitHub, each incorporating at least one LLM component, and then asked the following three research questions (RQs):

\noindent \textbf{RQ1: What are the characteristics of LLM-enabled apps?}

\noindent \underline{Method}: For each Android app, we leveraged the GitHub API~\cite{githubapi} to gather metrics such as the number of stars, commits, issues, and contributors. We then manually inspected each app to collect the LLM-related information, such as the types and their vendors. From this data, we summarized common traits across the dataset as well as outliers.

\noindent \underline{Results:} 
We analyzed and interpreted the distribution of the metrics above, finding that 
there is little difference in value between the first 2 quartiles, indicating that a large percentage of apps are smaller projects, but there are severe outliers. For example, the median star count is 8, but the most starred app in the dataset has 58,129. 
Finally, we classify the app functionalities across the dataset and discover that the majority of apps' main functionality is chatbots, followed by other uses of AI APIs such as text and image generation.


\noindent \textbf{RQ2: What main integration strategies are used to adopt LLMs into apps?}

\noindent \underline{Method}: We downloaded and qualitatively  analyzed the source code of each app   to identify the different methods used by developers to integrate LLMs into apps. We also provided descriptive statistics for releases, issues, LLM-related releases, and the app sizes associated with each strategy.

\noindent \underline{Results:} We identified five integration strategies (IS) that are utilized across the dataset to call LLMs as described below. We find that 132 out of 149 apps integrate LLMs through the first strategy, and apps that download the LLM either on the end-device or a back-end server require fewer updates to LLM-related code, compared to strategies that do not. 

\begin{itemize}
    \item \textbf{IS1:}  Using third-party APIs provided by LLM vendors.

  \item \textbf{IS2:}  Using a third-party website and calling LLMs as a webview app.

  \item \textbf{IS3:} Calling LLMs hosted on a back-end server.

  \item \textbf{IS4:}  Hosting LLMs on the user end-device.

  \item \textbf{IS5:} A hybrid strategy that utilizes two out of the four integration strategies.
\end{itemize}

\noindent

\noindent 
\textbf{RQ3: What drives developers to update LLM-related code?}

\noindent 
\underline{Method}: We manually identified the code used to interact with LLMs for each app then collected the number of commits to these code files as well as the commit messages. In order to understand the intention behind these updates, we fed the scrambled set of all commit messages into ChatGPT~\cite{chatgpt} to identify the top ten most common topics across these updates. In addition, we also observe the distribution of topics across each integration strategy.

\noindent \underline{Results:}
We observe that the top three most common topics for LLM-related code updates are \textit{Feature Additions}, \textit{Refactoring and Code Cleanup}, and \textit{Version Updates}. 


To sum up, our study makes the following contributions:
(1) We present a novel dataset of Android apps with LLM integration,
(2) We summarize the characteristics of this dataset and the strategies used to integrate LLMs into the apps, and
(3) We analyze the commit messages of LLM-related code updates to obtain insight into what drives developers to update LLM-related code. 
Our findings shed light on how LLMs have thus far been applied to mobile apps, and the challenges and opportunities that come with it. The entire dataset can be found in the replication package~\cite{rep-pkg}.


\section{Related Work}
\label{related-work}

\subsection{Studies on ML-enabled Systems}

With the popularity of machine learning (ML) increasing over recent years, there has been an increasing number of products incorporating ML models into their products~\cite{sens2024large,minaee2024large,zhao2023survey, meyer2023chatgpt, fan2023large, yang2024if}. However, adding ML models to existing systems require additional infrastructure and consideration of system architecture~\cite{paguthaniyaintegration, fan2023large}. As well, developers must consider system requirements and user concerns and experience, which poses challenges on how to best embed the ML model~\cite{sens2024large, nahar2023dataset}. The most commonly named problems include data quality, meshing ML models with traditional software parts, tool support for managing ML products, and quality assurance~\cite{sens2024large, paguthaniyaintegration}. Overall, we still know very little about how ML models are embedded and integrated into ML-enabled systems~\cite{sens2024large}. 

As a new subset of ML models, there is a similar mystery surrounding how LLM models are being integrated into LLM-enabled systems. Also, the findings from prior work focussing on general systems may not be directly applicable to mobile apps, as mobile apps are subject to more continuous, user-motivated, maintenance~\cite{viennot2014measurement}, and different non-functional constraints (e.g. energy consumption)~\cite{palomba2019impact}. We aim to focus on a subset of LLM-enabled systems by presenting an empirical study on 149 LLM-enabled Android apps and the strategies used to integrate LLMs. 



\subsection{On-Device LLM Deployment}

One method to integrate LLMs into a system is to deploy them on-device. Deploying LLMs on device is becoming a research hotspot as it increases the potential applications and eliminates the costs of cloud deployment, such as security and internet connection~\cite{naveed2023comprehensive}. However, there are restrictions on  limited hardware performance, memory bandwidth, and storage capacity. LLMs have many parameters and have large memory requirements to store the model parameters, the model activations, and the gradients and corresponding statistics~\cite{naveed2023comprehensive}.


To facilitate on-device deployment, LLM-specific inference engines have been developed, such as  Llama.cpp~\cite{llamacpp}, MLC-LLM~\cite{mlcllm}, or Langchain~\cite{langchain}. These engines are specifically designed for transformer-based LLM deployment on CPUs and GPUs. 
In this study, we aim to investigate the use of these engines in implementing LLMs in mobile apps, and how they are incorporated into the rest of the system.



\subsection{Third-Party Libraries in Mobile Apps}
A large number of companies develop software by way of Application Programming Interfaces (APIs), also known as third-party libraries, which allow the reuse of existing code components~\cite{wang2017understanding}. These libraries are widely used in mobile apps, accounting for, on average, over 60\% of Android app code as well as being used in almost every popular open-source Android app project~\cite{wang2017understanding, polese2022adoption}. When these companies issue updates to these libraries, app developers have the choice to adopt these implementations, but may be deterred due to the effort or effects of upgrading these libraries ~\cite{ahasanuzzaman2020studying}. 
Salza et al. conducted an empirical investigation into the when, why, and how mobile app developers update third-party libraries in code and found that only 15.52\% of library uses are constantly updated by developers, most of the updates are done with the aim of avoiding bug propagation or making an app compatible with the Android releases, and some app developers do not update libraries due to low payoff or to not break existing code~\cite{salza2020third}. We aim to investigate the particular usage of LLM-related APIs, and whether they are a large factor in-app updates and new releases.

Considering the numerous factors contributing to the uncertainty and challenges associated with implementing LLMs—including the requisite architectures, infrastructures, and best practices—we aim to address key research gaps through an analysis of Android apps that have integrated LLMs. This study seeks to identify prevalent use cases, challenges, and integration strategies related to incorporating LLMs, as well as to characterize the features of such apps. By leveraging these findings and metrics, we endeavor to answer critical research questions regarding the implementation of LLMs in Android apps, ultimately offering insights to inform and facilitate the integration of LLMs in mobile apps.

\section{Data Collection}
\label{data-collection}

\begin{figure}[t]
\centerline{\includegraphics[width=0.5\textwidth]{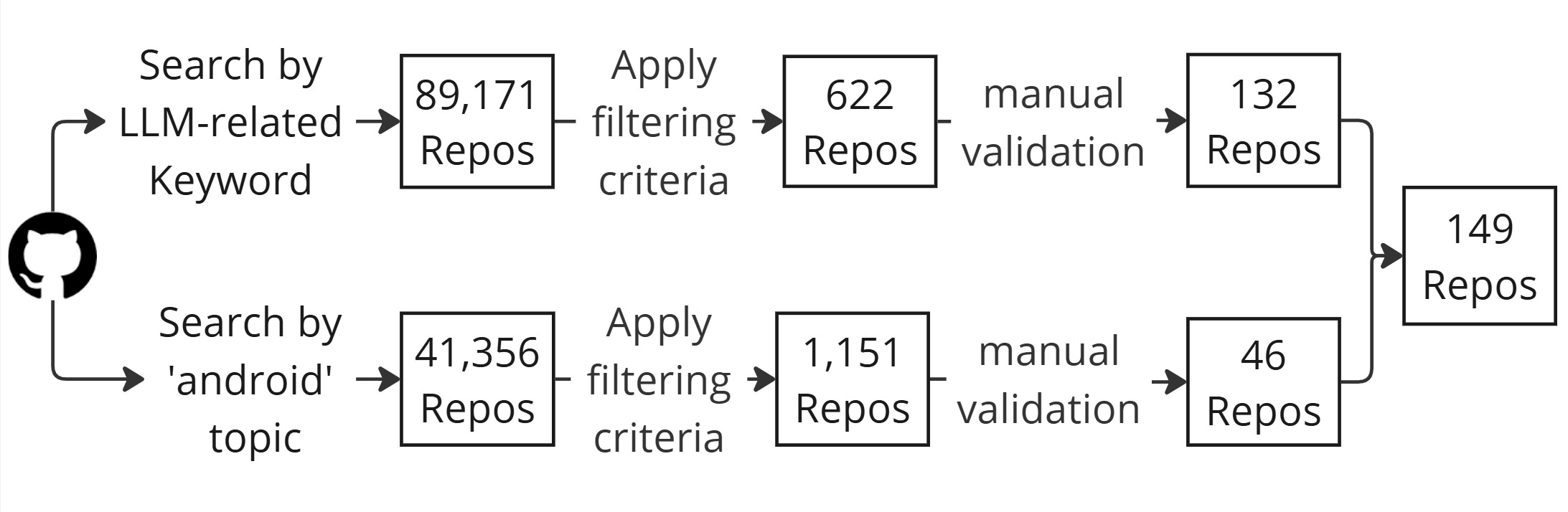}}
\vspace{-3mm}
\caption{Data collection overview.}
\label{fig-overview}
\vspace{-1.5em}
\end{figure}

Our study focused exclusively on open-source Android apps with LLM integrations. Open-source projects provide access to development practices typically unavailable in closed-source software, such as code architecture, commit history, and testing procedures. We began with a list of 33 LLM-related keywords obtained from model repositories and scholarly articles, covering widely-used LLMs, frameworks specific to LLMs, and companies focused on LLM technology~\cite{Ollama, hugginfaceleaderboard, novitaai, snitika, atasma, lben, gharry}, listed in Tab.~\ref{keyword-list}.
We utilized the GitHub API~\cite{githubapi} to collect a list of repositories based on two distinct search criteria as presented in Fig.~\ref{fig-overview}.

    \begin{table}[h]
    \vspace{-2mm}
    \caption{The LLM-related keywords used for data collection. }
\centering
\begin{tabularx}{\linewidth}{l>{\raggedright\arraybackslash}X}
\toprule
\textbf{Category}&{\textbf{Keywords for Data Collection }} \\
\midrule
Company& openai~\cite{openai}, cohere~\cite{cohere}\\
\midrule
Framework& langchain~\cite{langchain}, mlc llm~\cite{mlcllm}, ollama~\cite{Ollama}, llama~\cite{llamacpp}\\
\midrule
Term& LLM, large language model\\
\midrule
Model& alexatm~\cite{alexatm},  claude~\cite{claude}, deepseek~\cite{deepseek}, falcon~\cite{falcon}, gemini~\cite{gemini}, gemma~\cite{gemma}, gpt~\cite{chatgpt}, LaMDA~\cite{lamda}, marcoroni~\cite{marcoroni}, minichat~\cite{minichat}, mistral~\cite{mistral}, mixtral~\cite{mixtral}, nyxene~\cite{nyxene}, orca~\cite{orca}, PaLM~\cite{palm}, phi-1.5~\cite{phi1.5}, phi-3~\cite{phi3}, qwen~\cite{qwen}, redpajama~\cite{redpajama}, shining valiant~\cite{shiningvaliant}, stable beluga~\cite{stablebeluga}, stablelm~\cite{stablelm}, starling~\cite{starling}, una-xaberius~\cite{una-xaberius}, vicuna~\cite{vicuna}\\
\bottomrule
\end{tabularx}
\label{keyword-list}
\vspace{-0.5em}
\end{table}

\subsubsection{Search criterion 1: Matching keywords in repository names, descriptions, or issue/PR discussions}
We collected all repositories that had one or more of the keywords in their name or description, with at least one star and one fork. 
We also collected all repositories with one or more stars and forks with at least one issue or pull request (PR) that contained one or more of the keywords. We specified the creation date of the issues or PRs to be after the release date of the corresponding LLM, or within the year 2022 for LLM-related terms. This ensures that the search results focus on discussions, updates, or contributions that are relevant to the deployment or utilization of the LLM in question, reflecting more current and contextually appropriate data. This resulted in a list of 89,170 unique repositories.

Following the collection of repositories, we filtered repositories that match the following exclusion criteria (EC):

\begin{itemize}
    \item \textbf{EC1: Projects without a Readme file:}  We excluded all repositories that did not have a ReadMe file. This was performed to remove toy projects, as well as to prevent toy projects that are not well maintained from evading the next step. This reduced the number of repositories to 81,955.

    \item\textbf{EC2: Projects that are either deprecated or not Android apps:} We discarded all repositories that had at least one of the exclusion keywords listed in Tab.~\ref{exclusion-list} present in the ReadMe file. 
This was done to remove repositories that were deprecated, archived, or obvious non-product repositories, using exclusion terms such as ``tutorial,” ``framework,” and ``obsolete.” 
These exclusion keywords were taken from Nahar et al.'s work~\cite{nahar2023dataset}, which utilized a similar filtering process for extracting ML products. 
24,144 repositories were remaining following this step.

    \item\textbf{EC3: Projects that do not have an AndroidManifest.xml or lack activities in the AndroidManifest.xml file:} 
App developers need to include \texttt{AndroidManifest.xml} file that specifies the main settings of the app, such as required permissions and the minimum Android API version. In our study, repositories lacking an AndroidManifest.xml file were excluded, as its absence indicates the repository does not represent a genuine Android app. This filtering reduced the number of repositories to 670. Subsequently, repositories containing one or more \texttt{AndroidManifest.xml} files were further examined to ensure at least one activity was defined in the manifest, resulting in the exclusion of 48 repositories and leaving a total of 622.

\end{itemize}

 \begin{table}[h]
 \vspace{-2mm}
\caption{The exclusion keywords used for filtering repositories.}
\centering
\begin{tabularx}{\linewidth}{>{\raggedright\arraybackslash}X}
\hline
\texttt{deprecated, obsolete, framework, library, testing, toolkit, example, sample, guideline, guide, tutorial, blog, book, libraries, toolchain, interview notes, curated collection} \\
\hline
\end{tabularx}
\label{exclusion-list}
\vspace{-0.5em}
\end{table}

\noindent
\looseness=-1
\textbf{Manual validation.} Following these automatic filtrations, the 622 repositories remaining were subject to manual examination for LLM-relation. Manual validation was employed for high accuracy in the end dataset. 
Each repository was examined for some LLM involvement, such as the use of REST APIs for closed-source LLMs, LLM-specific frameworks, or discussion of LLM use in issues or pull requests. Repositories were discarded for having no LLM-use or being archived. The most common reason for repositories being discarded was being flagged for LLM-related keywords that did not refer to LLMs in the context that it was flagged for. This was most common for ubiquitous terms such as \textit{‘palm’}, \textit{‘falcon’} and \textit{‘orca’}, as well as \textit{‘large language model’}, as the GitHub API did not maintain the order of terms. Some repositories were also flagged for keywords such as \textit{‘GPT’} and \textit{‘Claude’} from issues, in which users asked or referenced GPT when having questions or writing code. A few projects were categorised as ‘unknown’ for whether they had LLM use, typically due to being mono-repositories or containing multiple projects, which contained both LLM-use and Android apps, but had unclear overlap. 

After categorising each repository, we filtered all non-English repositories. This resulted in 518 English repositories from the previous 622. Of the 518 English repositories, 132 were true positives, 340 were false positives, and 46 were categorised as ‘unknown’ or ‘other’.

\subsubsection{Search criterion 2: Searching for Android repositories}
We performed a second data collection process to collect repositories that did not have LLM mention in their title issues/PRs but used LLMs in their code. In this process, we used the GitHub API to gather all repositories that had been added to the \textit{'android'} topic on GitHub, with one or more stars and forks. Following this, we cloned each repository and applied our exclusion criteria (EC1 - EC3). This step reduced our pool of candidates from 41,356 repositories to 12,063. We then searched each repository for the presence of any of our LLM-related keywords in the code or title, which further reduced our candidates to 6,678 repositories. Due to the commonality of our keywords, we applied another filtration step from this pool by searching each repository for the presence of any keyword, preceded by either a space or a special character. The result of this step was 1,151 repositories, which we performed manual validation on to find 46 repositories with LLM use, of which 17 were not also found in the previous process. This brought our final dataset to 149 repositories. 
From this dataset, we performed an initial data exploration into project characteristics.


\begin{figure}[t]
\vspace{-2mm}
\centerline{\includegraphics[width=0.5\textwidth]{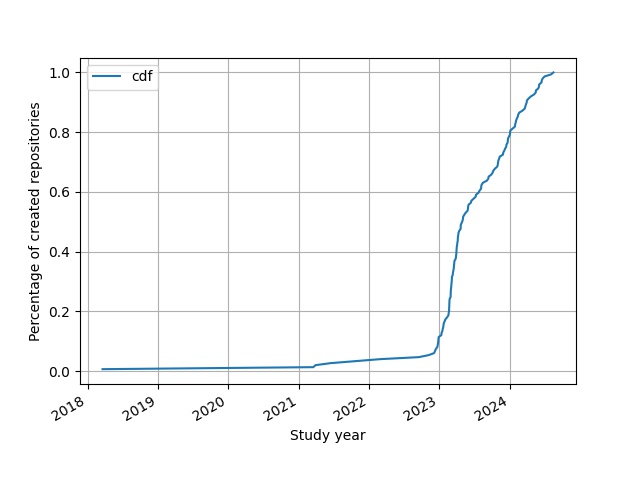}}
\vspace{-7mm}
\caption{Cumulative distributive function of creation time across the dataset of 149 Android apps.}
\label{cdf}
\vspace{-1em}
\end{figure}

\begin{table}[t]
\caption{Vendor, LLMs per vendor, and the number of apps per vendor.}
\centering
\begin{tabularx}{\linewidth}{l>{\raggedright\arraybackslash}Xc}
\toprule
\textbf{Vendor}&{\textbf{LLMs}} &{\textbf{ \# of apps}} \\
\midrule
Google &
Gemini, PaLM, flan-t5 &
41 \\ \midrule
Anthropic &
Claude &
5 \\ \midrule
Meta &
Llama, BART &
6 \\ \midrule
Openai &
GPT, Whisper, Dall-E &
111 \\ \midrule
Mistral &
mistral, mixtral &
7 \\ \midrule
Perplexity AI &
perplexity &
1 \\ \midrule
Ollama &
ollama &
1 \\ \midrule
Huggingface &
uggingchat &
1 \\ \midrule
Open-source vendors  &
local model  &
4 \\

\bottomrule
\end{tabularx}
\label{vendors-llms-count}
\vspace{-2.71em}
    \end{table}

\section{RQ1: What are the characteristics of LLM-related apps?}
\label{rq1}

\subsection{Methodology}


To determine the characteristics of the dataset, we gathered repository metadata related to the popularity and relevance of projects, including the counts for contributors, commits, and issues for each app using the GitHub API ~\cite{githubapi}. Additionally, we used cloc~\cite{cloc} to count blank lines, comment lines, and physical lines of source code in a code base, to determine the software size in lines of code (LOC) per app. 
We then manually investigated each app to log which LLMs were used per app, whether it was open- or closed-source, as well as used locally or with an API. 

\looseness=-1
Finally, we classify the functionalities offered by the apps of this dataset. For the 132 of 149 apps that had a repository description, we retrieved the descriptions used and submitted them to OpenAI’s GPT-
4o-mini using a paid OpenAI API subscription, set to zero temperature~\cite{chatgpt}. We first asked the model to determine the top 5 functionalities presented across all repository descriptions in a zero-shot prompt and to provide a description of each functionality. We then prompted the model with another zero-shot prompt that provided the list of repository descriptions and the top 5 functionalities and their descriptions and asked the model to assign one functionality to each app. Finally, for the 17 apps that did not have a repository description, we manually categorized them by title and ReadMe file. The results of these prompts are shown in Tab.~\ref{app-functionalities}.


\begin{figure*} [t]
\label{histograms}
\includegraphics[width=.23\linewidth]{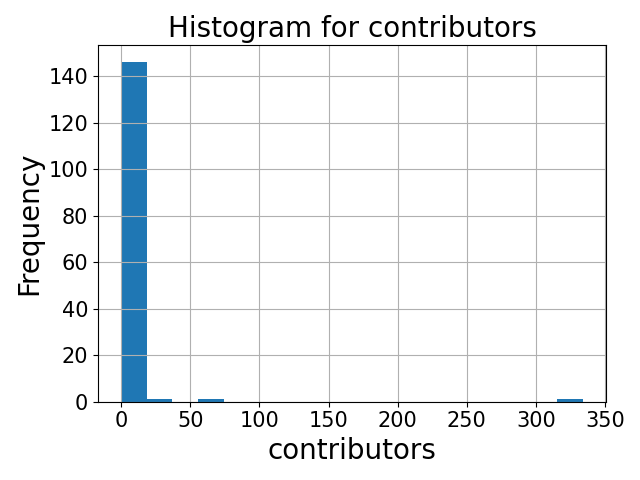}
\includegraphics[width=.23\linewidth]{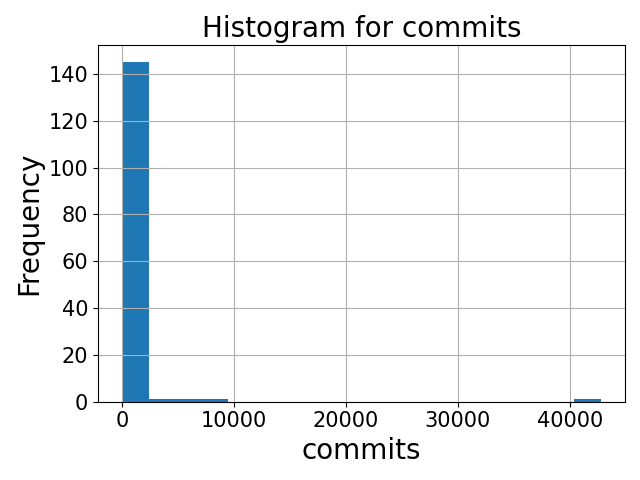}
\includegraphics[width=.23\linewidth]{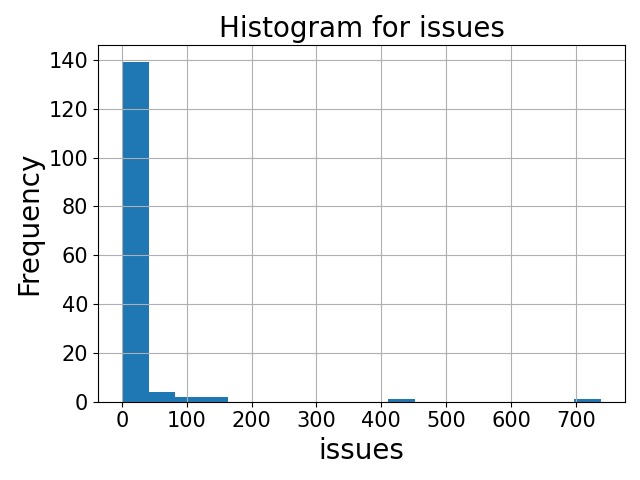}
\includegraphics[width=.23\linewidth]{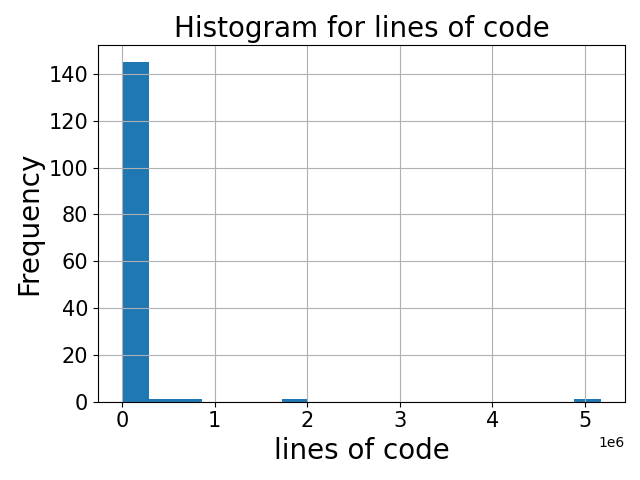}
\vspace{-3mm}
\caption{Histograms displaying the distribution of several characteristics.}
\vspace{-1.5em}
\end{figure*}

\subsection{Results}

The distribution for the collected metrics is shown in Fig.~\ref{histograms} and also presented in Tab.~\ref{summary-statistics}.
Fig.~\ref{cdf} shows the cumulative distribution of the creation date of repositories, with the earliest repository (\textit{``YoubiMiku"}) being created in 2018.
We also present the top 10 repositories by star count in Tab.~\ref{top_ten_repos}.
We also observe a large range of values for the number of commits (spanning from 19 to 42,801) and contributors (1 to 334) of these 10 apps. 
Notably, six out of the 10 apps are also available on Google Play Store.
Tab.~\ref{vendors-llms-count} displays the overall distribution of LLMs used in this dataset, and their vendors. 

\begin{table}[t]
\caption{An overview of the top ten repositories sorted by star count.}\label{top_ten_repos}
\resizebox{\columnwidth}{!}{\begin{tabular}{lccccc}
\toprule
\textbf{App}& \textbf{\#Stars} &{\textbf{\#Commits}}&{\textbf{\#Contributors}}&{\textbf{\#Issues}}&{\textbf{Google Play}} \\
\midrule
\href{https://github.com/AppFlowy-IO/AppFlowy}{AppFlowy} &
58,129 &
6,294 &
334 &
436 &
\href{https://play.google.com/store/apps/details?id=io.appflowy.appflowy}{Yes} \\
\midrule
\href{https://github.com/Mobile-Artificial-Intelligence/maid}{Maid}&
1,485 &
1,885 &
22 &
158 &
\href{https://play.google.com/store/apps/details?id=com.danemadsen.maid}{Yes} \\
\midrule
\href{https://github.com/futo-org/android-keyboard}{Futo Keyboard}&
775 &
42,801 &
62 &
739  &
\href{https://play.google.com/store/apps/details?id=org.futo.inputmethod.latin.playstore}{Yes} \\
\midrule
\href{https://github.com/Skythinker616/gpt-assistant-android}{GPT Assistant}&
667 &
44 &
1 &
51  &
No \\
\midrule
\href{https://github.com/flyun/chatAir}{chatAir}&
527 &
573 &
15 &
47  &
\href{https://play.google.com/store/apps/details?id=info.flyun.chatair}{Yes} \\
\midrule
\href{https://github.com/wewehao/flutter_chatgpt}{Flutter ChatGPT} &
461 &
19 &
6 &
9   &
No \\
\midrule
\href{https://github.com/jakepurple13/OtakuWorld}{OtakuWorld} &
446 &
2,176 &
5 &
11  &
No \\
\midrule
\href{https://github.com/Taewan-P/gpt_mobile}{GPT Mobile} &
413 &
377 &
4 &
38  &
\href{https://play.google.com/store/apps/details?id=dev.chungjungsoo.gptmobile}{Yes} \\
\hline
\href{https://github.com/matthaigh27/ChatGPT-android-app}{ChatGPT App} &
337 &
19 &
2 &
22 &
No \\
\midrule
\href{https://github.com/AndraxDev/speak-gpt}{SpeakGPT} &
291 &
179 &
2 &
128   &
\href{https://play.google.com/store/apps/details?id=org.teslasoft.assistant}{Yes} \\
\bottomrule
\end{tabular}}
\vspace{-2em}
\end{table}

\textbf{Contributor Distribution.}
Our results show that the average number of contributors is 5 with a median of 1 contributor.
The minimum, first quartile, and second quartile values for the number of contributors are all 1, indicating that at least half of the apps are maintained by a single contributor. 
The third quartile shows only a modest increase, reaching 2 contributors. However, the dataset contains a significant outlier: the app \textit{``AppFlowy''}~\cite{appflowy}, which has 334 contributors, far exceeding the third quartile value.
As described in the project Readme file, this app offers a variety of LLMs via APIs, including GPT, Claude, Llama, and Mistral, as well as locally hosted models, to power their search functionality.

In contrast, the majority of apps (101 out of 149) have only 1 contributor. These apps exhibit limited activity, with median values of 0 issues, 0 releases, 2 stars, and 34 commits. The remaining 48 apps with more than one contributor demonstrate higher engagement with a median of 21.5 stars and 21.5 commits, though their median number of issues and releases remains the same as those maintained by a single contributor.

\textbf{Commit size.}
Although the average number of commits is 501, only 25\% of the projects have more than 60 commits, and 50\% have fewer than 20 commits. 
There is an average of 7 days per commit across all apps, with a median of 1 day per commit.
Notably, \textit{``Futo Keyboard''} app~\cite{futokeyboard} stands out with 42,801 commits, significantly exceeding the dataset's general trends. \textit{``Futo Keyboard''} is a customizable keyboard app leveraging local LLMs Whisper and Llama, and is among the more popular apps, with 775 stars and 62 contributors.
The minimum number of commits is recorded for the app \textit{``WizGPT''}~\cite{WizGPT}, which was created on May 4, 2023, with only one commit labeled as ``initial commit" since then.
\textit{``WizGPT''} was planned to be a GPT and Dall-E powered voice assistant that can be run on Android, iOS, Web, and Desktop based on the Readme file. 
\textit{``WizGPT''} has  6 stars and 0 releases and issues.  

\textbf{Number of issues.} 
The dataset shows a median of 0 issues per project, with an average of 16 issues. Half of the projects have no reported issues, and three-quarters have 4 or fewer. The app \textit{``Futo Keyboard''}~\cite{futokeyboard} significantly skews the upper range with 739 issues, which inflates the average despite most projects exhibiting minimal or no issues.
A total of 82 apps, representing the majority, have 0 issues. These projects typically exhibit minimal activity, with median values of 0 releases, 4 stars, and 11 commits. In contrast, apps with at least 1 issue demonstrate higher engagement, with median values of 1 release, 24 stars, and 38 commits.

\textbf{Sizes of the apps.} 
The median size of the collected apps, measured in LOC, is 3,838, with an average size of 70,360 LOC. The average LOC is nearly twenty times the median, indicating a skewed distribution with some exceptionally large apps.
The app with the highest LOC, at 5,176,508, is \textit{chatAir}~\cite{chatair}, a native Android app for chatting with GPT, Claude, and Gemini. As one of the most prominent apps in the dataset, \textit{chatAir} boasts 485 stars and ranks above the upper quartile for issues, contributors, and commits. Its LOC count is over five times the dataset's average and represents the largest size among all projects. The app also has the most tags in the dataset, reflecting its active development.
Conversely, the app with the smallest LOC is \textit{OldGPT}~\cite{oldgpt}, a lightweight chat app designed for users with outdated browsers or operating systems. \textit{OldGPT} integrates GPT through the OpenAI API and has minimal activity, with just 1 star, 3 releases, and no reported issues.

\textbf{App functionalities.}
As shown in Tab.~\ref{app-functionalities}, the most common functionality seen across the repository descriptions is Chatbot. Over half of the dataset (56\%) is categorized into this functionality. 
Conversely, the least common functionality is Task Management and Productivity Tools, with only 6 apps categorized into this functionality. Examining the descriptions of each functionality, we can see that all functionalities mention AI, indicating that the majority of apps use LLMs to provide a large part of their service.



\begin{table}[t]
 \footnotesize
\caption{A summary of the distribution of quantitative measurements taken on the dataset.Q1--First Quartile, Q3--Third Quartile.}
\label{summary-statistics}
\begin{tabular}{lcccccc}
\toprule
\textbf{Category}&{\textbf{Min.}}&{\textbf{Q1}}&{\textbf{Median}}&{\textbf{Q3}}&{\textbf{Max.}}&{\textbf{Mean}} \\
\midrule
\#Stars  &
1	& 3 &	8 &	27	& 58,129	& 449.4 \\\midrule

\#Commits  &
 1 &	7 &	22 &	61.0&	42,801 &	500.7 \\
\midrule
\#Releases  &
0	& 0	& 0 &	1 &	88 &	4.6 \\
\midrule
Project age (days) & 0
 & 4 & 64 & 293 & 2,432 & 199 \\
\midrule
\#Contributors &
1	& 1	&1	&2	& 334 &	4.7 \\
\midrule
\#Issues &
0 &	0	& 0	& 4 &	739	&15.7 \\
\midrule
KLOC&
0.2 &	2.2&	3.8	& 9.1 &	5176.5	& 70.3 \\
\midrule
Avg. days/commit&
0.0	 & 0.3 &	1.3	&4.5 &	195.3	& 6.8 \\
\bottomrule

\end{tabular}
\vspace{-1em}
\end{table}

\begin{table*}[t]
\caption{App functionalities across the dataset.}
\label{app-functionalities}
\centering
\begin{tabular}{ L{1.9cm} c L{9.5cm}  L{4.3cm}  }
\toprule
\textbf{Functionality}&{\textbf{\# (\%)}}&{\textbf{Description}}&{\textbf{E.g., App and Description}} \\
\midrule
Chatbot & 84 (56\%) & Apps feature a chatbot interface that allows users to engage in natural language conversations with AI models like ChatGPT or Gemini. This functionality enables users to ask questions, seek information, or have casual conversations, providing a personalized and interactive experience. & \href{https://github.com/chouaibMo/ChatGemini}{ChatGemini} - this app is a multiplatform chatbot app powered by Gemini. \\

\midrule
Integration with LLM APIs & 27 (18\%) & Apps utilize APIs from OpenAI or other AI providers to access advanced features such as text generation, image creation, and summarization. This integration allows developers to leverage powerful AI models to enhance the app's capabilities and provide users with sophisticated functionalities. & \href{https://github.com/F0x1d/Sense}{Sense} - this app's description describes itself as an OpenAI client with ChatGPT support.
\\
\midrule
Content Generation & 19 (13\%) & Apps are designed to generate content based on user input, such as recipes, stories, or summaries of videos and documents. This functionality allows users to create personalized content quickly and efficiently, often using AI to enhance creativity and reduce effort.  &  \href{https://github.com/mhss1/AIStudyAssistant}{AI Study Assistant} - this app aims to enhance learning experiences by offering lecture summarization, essay writing, and question generation.
\\
\midrule
Voice Interaction &
13 (9\%) & Apps incorporate voice recognition and synthesis capabilities, allowing users to interact with the AI using voice commands. This functionality enhances accessibility and convenience, enabling hands-free operation and real-time voice responses. & 
\href{https://github.com/AndraxDev/speak-gpt}{SpeakGPT} - this app is described as a personal voice assistant, based on ChatGPT.
\\

\midrule
Task Management and Productivity Tools &
6 (4\%) & Apps focus on productivity by offering features like task management, to-do lists, and journaling. These apps often use AI to help users generate tasks, track their mood, or summarize information, making it easier to manage daily activities and improve mental well-being. & 
\href{https://github.com/alexandresanlim/flutter-todo-list-chat-gpt}{ToDo list with ChatGPT} - this app's functionality is facilitating to-do lists by generating tasks using ChatGPT.
\\
\bottomrule
\end{tabular}
\end{table*}


\section{RQ2: What main integration strategies are used to adopt LLMs into apps?}
\label{rq2}

\subsection{Methodology}
\noindent
To identify the strategies employed for integrating LLMs within the dataset, we analyzed the source code for each app by extracting and inspecting files containing code responsible for importing or invoking LLM functionalities. Similar methods of model invocation were categorized into distinct integration strategies, for which we computed median metrics, including frequency of usage, number of releases, reported issues, and lines of code. Furthermore, we calculated the median proportion of LLM-related releases by pinpointing commits that modified files containing LLM-related code. We then counted the number of releases that followed one or more such commits since the prior release. LLM-related code was defined as code-facilitating tasks such as selecting, switching, downloading, or invoking LLMs, including elements such as API URLs, model identifiers, or prompt templates.


\subsection{Results}
\noindent
Through qualitative analysis, we identified five distinct strategies employed by developers to integrate LLMs into their workflows. Tab.~\ref{Strategy_Statistics} provides an overview of the key characteristics associated with each strategy. These strategies are described in detail below.

\begin{table*}[t]
\caption{A summary of the distribution of quantitative measurements by integration strategy.}
\label{Strategy_Statistics}
\centering
\begin{tabular}{lccccc}
\toprule
\textbf{Integration Strategy}&{\textbf{Frequency}}&{\textbf{Median \#releases}}&{\textbf{Median \#issues}}&{\textbf{Median \%LLM-related releases}}&{\textbf{Median \#LOC}} \\
\midrule
IS1: Using third-party API calls & 132 & 0.0 & 0.0 & 66.7\% & 2,929.0 \\

\midrule

IS2: Using third-party website & 6 & 6.5 & 6.5 & 83.3\% & 916.5 \\

\midrule

IS3: Hosting LLMs on back-end servers & 5 & 1.0 & 1.0 & 0.0\% & 6,519.0 \\
\midrule

IS4: Hosting LLMs on user devices & 2 & 4.0 & 371.5 & 7.1\% & 26,3723.5 \\
\midrule

IS5: Mixing multiple integration strategies & 4 & 21.0 & 100.5 & 40.3\% & 61,890.5 \\

\bottomrule

\end{tabular}
\end{table*}

\medskip
\subsubsection{\textbf{Strategy 1: Using Third-party API Calls}}
This strategy is by far the most common across the dataset, with 132 apps in our dataset using this integration strategy.

\begin{figure}[t]
\centerline{\includegraphics[width=0.4\textwidth]{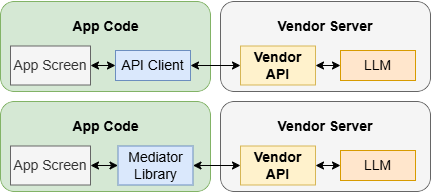}}
\caption{An overview of the two methods for calling LLMs using Third-party APIs.}
\label{Strat1_Architecture}
\vspace{-1.7em}
\end{figure}

\begin{figure}[htbp]
\vspace{-2mm}
\centerline{\includegraphics[width=0.4\textwidth]{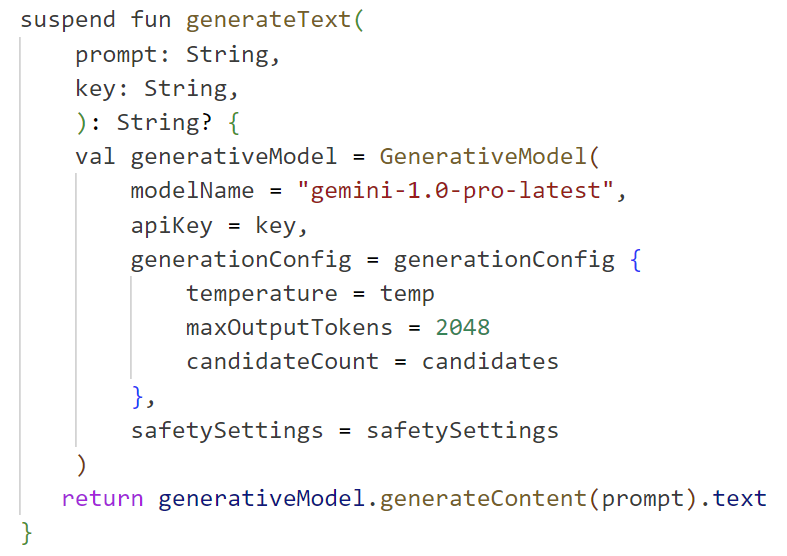}}
\vspace{-2mm}
\caption{An example of integrating Gemini using Google API~\cite{aistudyassistant}.}
\label{Strat1_Example}
\vspace{-0.5em}
\end{figure}

\noindent{\textbf{Description:}}
\noindent
This strategy (shown in Fig.~\ref{Strat1_Architecture}) allows developers to access popular closed-source LLMs (e.g., GPT and Gemini) using an API key.
As shown in Fig.~\ref{Strat1_Example}, developers set up the model parameters (e.g., select the used model, define the required temperature, and the maximum output token size).
Then, the mobile app calls the third-party model using an API key for authentication.
Given this strategy's popularity, packages are available from both vendors and third parties that can help implement it. 
Two-thirds (87) of this strategy's apps call directly the third-party API using HTTP requests. 

Our analysis reveals that 45 apps employ mediator libraries, which serve as intermediaries to facilitate integration with third-party APIs. These mediator libraries can be categorized into two distinct types:
\begin{enumerate}
    \item  \textbf{ Custom mediators that facilitate the connection to a single vendor's API:} In this type, the third-party API providers (e.g., Google, OpenAI) design the mediator library to facilitate the connection to their API.
For example, as shown in Fig.~\ref{Strat1_Example}, \textit{``AI Study Assistant''} app imports Google's official generative library to access the GenerativeModel object for calling Gemini. 
We find 21 apps that adopt custom mediators to integrate LLMs into their code.

  \item \textbf{ Generic mediators that facilitate the connection to multiple vendors' API:} In this type, a generic mediator library that facilitates the connection to multiple LLM APIs. 
For example, \textit{``ChatAir''} app imports the theokanning.openai package that facilitates a connection to GPT, Gemini, and Claude.
We found 24 apps that adopt generic mediators to integrate LLMs into their code.

\end{enumerate}
\noindent{\textbf{Insights and Discussion:}}
Consistent with the overall dataset, the subset of apps utilizing third-party API calls exhibits a right-skewed distribution across most metrics. The median values for both releases and issues in this subset are zero. However, for apps with at least one release, the median number of releases is 2.5, with 66.7\% of these releases involving modifications to LLM-related code. These findings highlight the necessity for developers to actively maintain and update LLM-related code as part of their app development process.


\subsubsection{\textbf{Strategy 2: Using Third-party Website}}
Six apps use a web view wrapper to access third-party LLMs. 

\begin{figure}[htbp]
\vspace{-2mm}
\centerline{\includegraphics[width=0.4\textwidth]{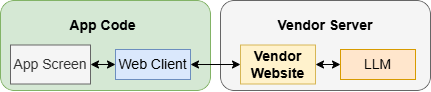}}
\caption{An overview of calling LLMs using a webview wrapper to submit prompts to the official vendor website.}
\label{Strat2_Architecture}
\vspace{-0.5em}
\end{figure}

\begin{figure}[htbp]
\centerline{\includegraphics[width=0.4\textwidth]{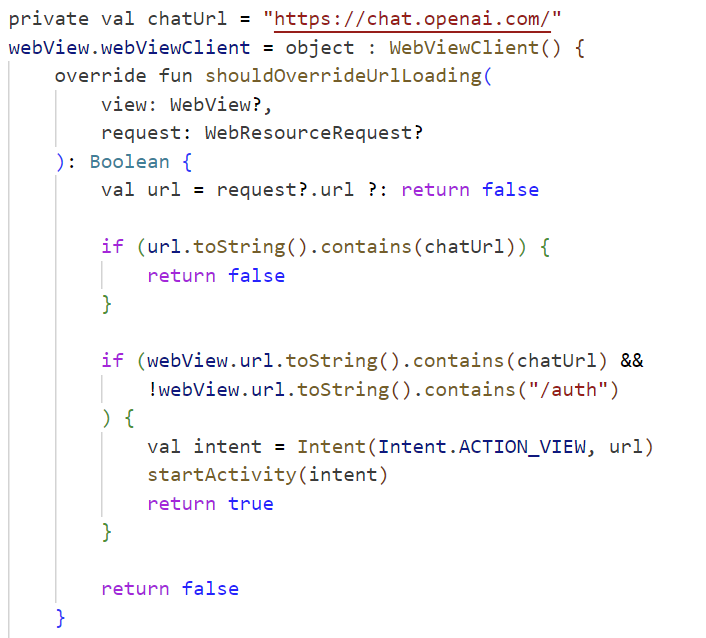}}
\vspace{0\baselineskip}
\caption{An example of using a webview client to send prompts with OpenAI's chat URL~\cite{ChatGPT-android-app}.}
\label{Strat2_Example}
\vspace{-0.5em}
\end{figure}

\noindent{\textbf{Description:}}
Vendors (e.g., ChatGPT~\cite{chatgpt} and HuggingChat~\cite{huggingchat}) can allow users to access their LLMs via their webpage. 
As shown in Fig.~\ref{Strat2_Architecture}, apps can facilitate sending prompts and receiving responses by implementing a webview client to load the Vendors' webpage URL. 
This strategy avoids calling an API to access closed-source LLMs (shown in Fig.~\ref{Strat2_Example}). 
However, it may offer reduced capabilities compared to those provided by APIs, due to rate limits placed by the vendor ~\cite{chatgptratelimits}.

\noindent{\textbf{Insights and Discussion:}}
We find that five apps access the ChatGPT website~\cite{chatgpt} and one app accesses HuggingChat~\cite{huggingchat}, a chat interface provided by Huggingface that offers a list of LLMs to chat with. 
This strategy had the highest median percentage of LLM-related releases (83.8\%). When examining the code updates and commit messages associated with these releases, we observed that developers frequently tweaked code surrounding the web client. This may be because this strategy is not an officially provided integration method, so developers may need to work around changes to the vendor websites, such as updating URLs or domains. 


\subsubsection{\textbf{Strategy 3: Hosting LLMs on Back-end Servers}}
Five apps utilize LLMs that are hosted on back-end servers.

\noindent{\textbf{Description:}}
In this integration strategy, developers either host the LLM on a back-end server and allow their apps to access the hosted LLM using API calls, or provide the infrastructure for the user to host the LLM on the user's server. 
This strategy avoids memory and computation demands on the user-end devices and reduces the risks of sending users's data to third-party vendors (like sending users' photos/text to OpenAI). 
However, this strategy requires app developers or the user to set up and run a live server on a separate device. 

\noindent{\textbf{Insights and Discussion:}}
Out of the five apps, two have no releases. The median number of releases is 1 but rises to 5 when excluding the two apps without releases. Of the three apps that have one or more releases, two have no LLM-related releases, resulting in a median of 0.0\% of LLM-related releases. These results show that hosting LLMs on back-end servers may require less maintenance compared to calling third-party APIs. 
This could be because using a local LLM does not require the developers to adjust to the updates from the vendor side.

\begin{figure}[htbp]
\vspace{-2mm}
\centerline{\includegraphics[width=0.36\textwidth]{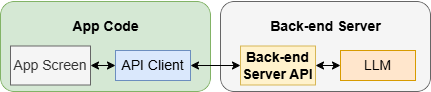}}
\caption{Calling LLMs hosted on a back-end server.}
\label{fig}
\end{figure}

\begin{figure}[htbp]
\centerline{\includegraphics[width=0.4\textwidth]{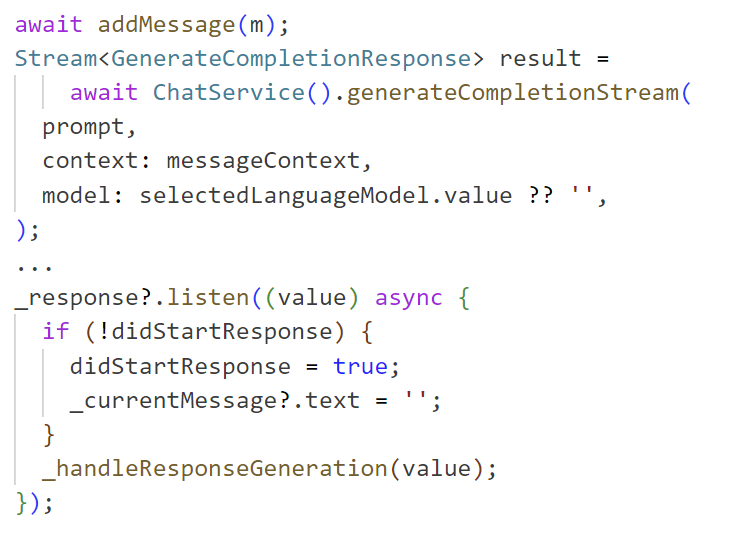}}
\caption{An example view model for receiving a response from an LLM hosted on a back-end server~\cite{amallo}.}
\label{fig}
\vspace{-0.5em}
\end{figure}

\begin{figure}[htbp]
\centerline{\includegraphics[width=0.5\textwidth]{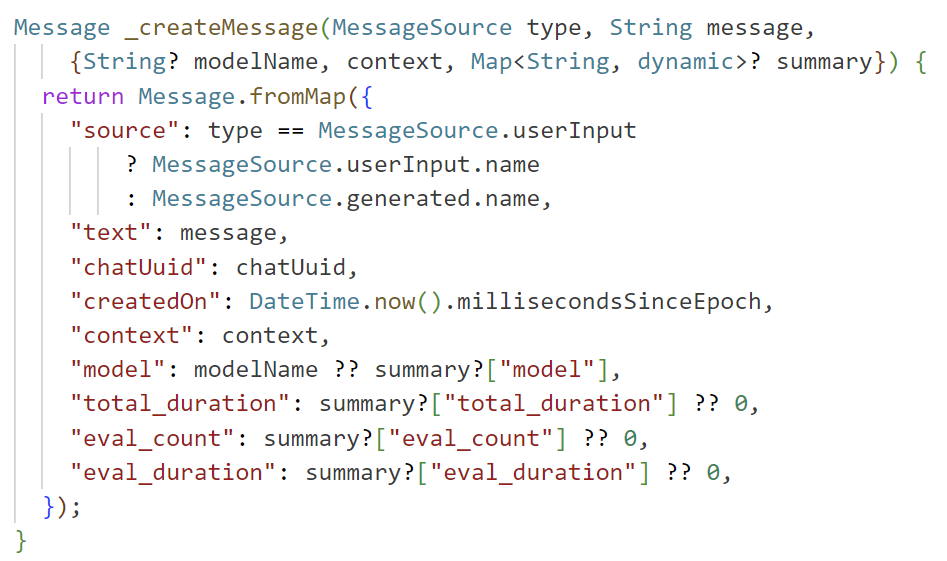}}
\caption{An example view model for creating a response from an LLM hosted on a back-end server~\cite{amallo}.}
\label{fig}
\vspace{-0.5em}
\end{figure}

\subsubsection{\textbf{Strategy 4: Hosting LLMs on User Devices}}
This is the least popular integration strategy across the dataset, as we find two apps in the dataset that used LLMs locally. 

\begin{figure}[htbp]
\vspace{-2mm}
\centerline{\includegraphics[width=0.3\textwidth]{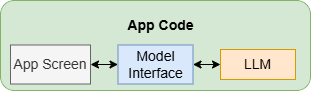}}
\caption{An overview of calling LLMs hosted on user devices.}
\label{Strat4_Architecture}
\vspace{-0.5em}
\end{figure}

\noindent{\textbf{Description:}}
This strategy (shown in Fig.~\ref{Strat4_Architecture}) requires the user to load and store the LLM on the end device, which consumes memory and computation but allows the user to run inference privately and without requiring the internet. 
However, we notice that app developers do not use frameworks to facilitate running LLMs locally, requiring developers to develop their own app-specific infrastructure.

\noindent{\textbf{Insights and Discussion:}}
Unlike apps that apply a hybrid strategy, which offers options for local models, both apps are built for specific LLMs. 
One of the two apps, \textit{``Futo Keyboard''}~\cite{futokeyboard} app, uses llama.cpp~\cite{llamacpp} and whisper.cpp~\cite{whispercpp} to run Llama and Whisper models with few parameters, reducing memory demands on the end device. 
The other runs an open-source GPT4All model. 

The \textit{``Futo Keyboard''} app also has the largest number of commits and issues in the dataset, which is the cause of this strategy's high median for its number of issues. Despite hosting the model and running inference locally, as well as having the highest median software lines of code, this strategy also has a relatively low median percentage for releases that contained updates to LLM-related code. Similar to the previous strategy, this could be because the local LLM requires fewer updates once implemented.

\begin{figure}[htbp]
\centerline{\includegraphics[width=0.5\textwidth]{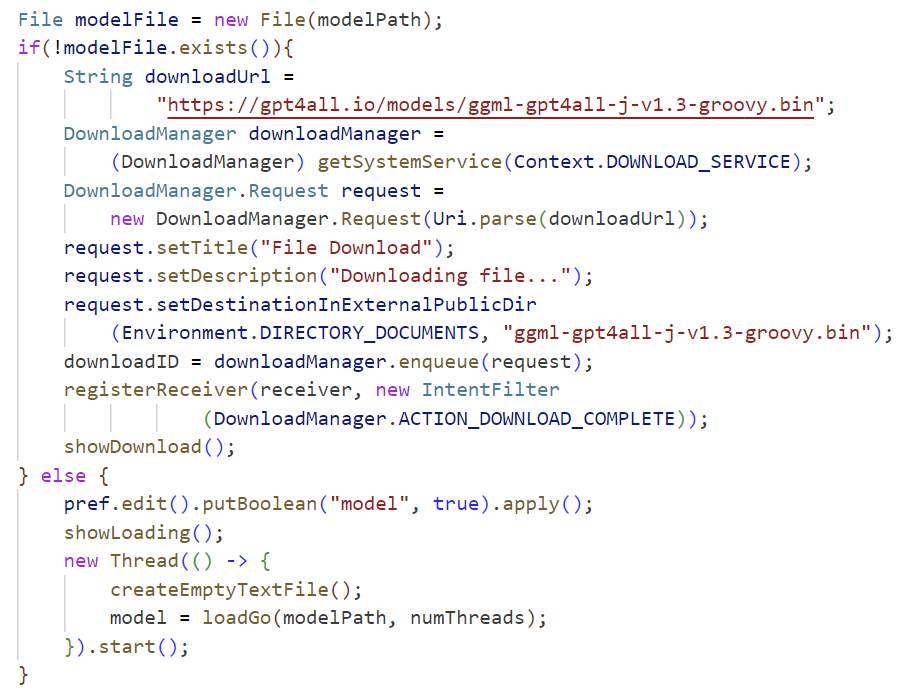}}
\vspace{-1mm}
\caption{An example of starting a chat with a local model~\cite{LocalGPT-Android}.}
\label{Strat4_Example}
\end{figure}

\subsubsection{\textbf{Strategy 5: Mixing Multiple Integration Strategies}}
There are four occurrences of a hybrid strategy in our dataset.

\noindent{\textbf{Description:}}
In this strategy, developers utilize multiple integration strategies, giving users choices between using vendor-provided APIs or calling LLMs locally or from a third party.

\noindent{\textbf{Insights and Discussion:}}
Two apps used LLMs both locally and through an API, and two apps used LLMs through both API and back-end servers. 
This subset contains some of the most popular apps in the dataset, with three out of the four apps in the third quartile across the entire dataset for stars and contributors, and all four in the third quartile for issues and releases. In particular, the \textit{``AppFlowy''} app, which supports APIs for GPT and Claude as well as local models such as Mistral and Llama, has the most stars and releases in the entire dataset, as detailed in RQ1. Another very popular app using this hybrid strategy is \textit{``Maid''} app~\cite{maid}, which is a chatting app that also offers LLMs both through API and locally. This app supports APIs for GPT, Claude, and Gemini, and local support for Ollama, Llama,  Mistral, and other local models, and is second in stars to \textit{``AppFlowy''} app.
As a result, this strategy has a high median number of releases and issues. The median percentage of LLM-related releases is about one in every four releases.



\section{RQ3: What drives developers to update LLM-related code?}
\label{rq3}

\subsection{Methodology}
\begin{figure}[t]
\centerline{\includegraphics[width=0.5\textwidth]{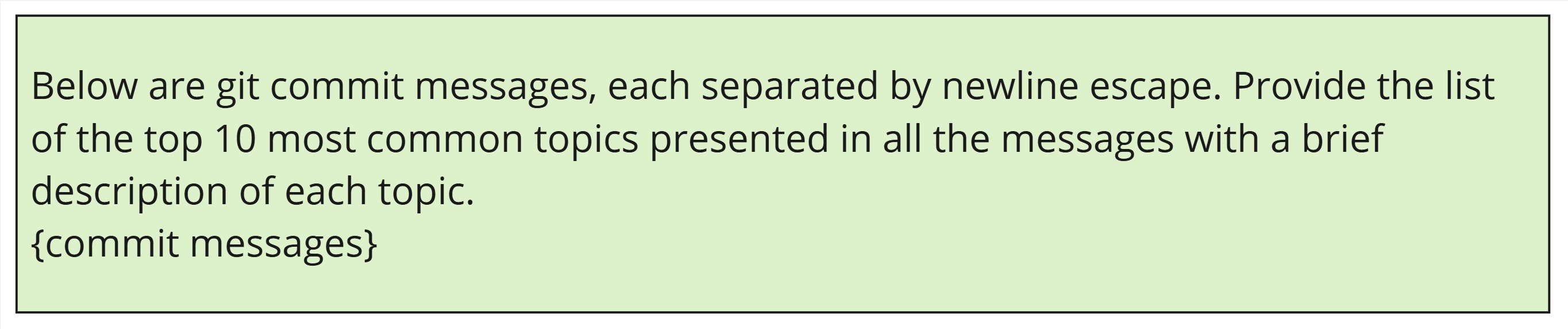}}
\vspace{-3mm}
\caption{Prompt to generate the ten most common topics.}
\label{Prompt1}
\vspace{-1.7em}
\end{figure}

To determine what drives developers to update
LLM-related code, we collected the commit messages from each LLM-related update across all repositories, totaling 2,003 commits, and submitted them to OpenAI's GPT-4o-mini using a paid OpenAI API subscription, set to zero temperature~\cite{openai}. 
We submitted two prompts to the model. The first was a zero-shot prompt that provided all of the commit messages, randomized, and asked the model to provide the list of the top ten most common topics across all of the commit messages as well as a description of each topic (shown in Fig.~\ref{Prompt1}). 

\begin{figure}[htbp]
\vspace{-2mm}
\centerline{\includegraphics[width=0.5\textwidth]{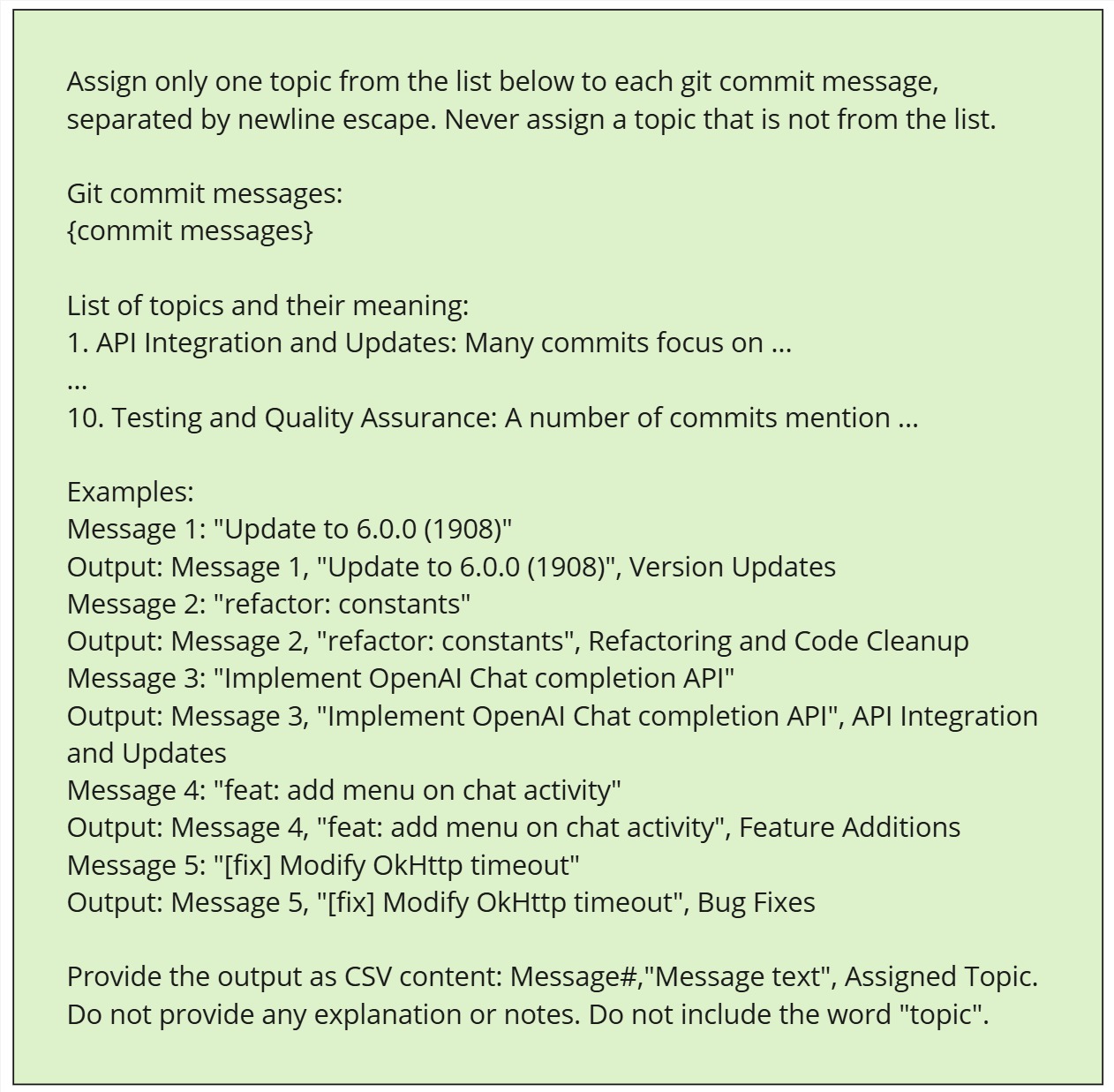}}
\vspace{-3.2mm}
\caption{Prompt to assign each commit message to a topic.}
\label{Prompt2}
\end{figure}

\looseness=-1
The second prompt (shown in Fig.~\ref{Prompt2}) is a few-shot prompt that provides the model with the commit messages in batches of 100 messages.
We inputted the list of topics and their descriptions, generated by the previous prompt, and five examples of commit messages and the expected outputs, and asked the model to assign each commit message to one of the topics. 

\subsection{Results}
\noindent
\textbf{We observe that \textit{Feature Additions}, \textit{Refactoring and Code Cleanup}, and \textit{Version Updates} are the most discussed topics when developers change LLM-related code.}
Tab.~\ref{commit_Message_Stats} shows the identified ten topics along with their description, frequency, and example commit messages. 
As shown in table~\ref{commit_Message_Stats}, the most common topics are\textit{``Feature Additions''}, \textit{``Refactoring and Code Cleanup''}, and \textit{``Version Updates''}. 
We also notice that \textit{``Bug fixes''} and \textit{``Error handling''} topics contribute to 242 commits (12.1\%), which indicates the prevalence of developers' mistakes while adopting LLMs in their apps. 
We observe that the least common topic is \textit{``Testing and Quality Assurance''}, which may indicate that developers need more testing effort to ensure that the integrated LLMs work properly. 

\begin{table*}[h]
\caption{The commit message topics.}
\label{commit_Message_Stats}
\centering
\begin{tabular}{ L{2.2cm}  r L{7.5cm}  L{5.4cm} }
\toprule
\textbf{Topic}&{\textbf{\# (\%)}}&{\textbf{Description}}&{\textbf{E.g., Commit Messages}} \\
\midrule

Feature Additions &
607 (30\%) & 
Messages indicate the addition of new features, such as voice input/output, image generation, and chat functionalities. &
\textit{``feat: suppport translate and summary using local ai (\#5858)''}
\\
\midrule

Refactoring and Code Cleanup & 
442 (22\%) & 
Messages indicate code refactoring activities, such as renaming variables, restructuring files, and removing unused code. &
\textit{``refactor: move openai state, store and logic into its own service''}
\\
\midrule

Version Updates & 
293 (15\%) &
Messages indicate updates to the app's version. &
\textit{``Upgrade LangChain.dart to v0.5.0''}
\\
\midrule

Bug Fixes & 
190 (10\%)  &
Messages indicate fixing bugs, such as handling crashes, incorrect API responses, and UI glitches.&
\textit{``fix: local ai enable\/disable (\#6151)''} 
\\
\midrule

API Integration and Updates  &
177 (9\%) &
Messages indicate the integration of new APIs or updating the integration of existing APIs, such as modifying code to ensure compatibility with the latest API specifications. &
\textit{``Fixed the model to the latest gpt-3.5-turbo model''} 
\\
\midrule

User Interface (UI) Improvements &
156 (8\%) & 
 Messages indicate enhancing the user interface, such as layout changes, theme updates, and overall design improvements. &
 \textit{``don't show cost in \$ but in tokens''}
\\
\midrule

Error Handling &
52 (3\%) &
Messages indicate improving error handling mechanisms, such as adding error messages and handling exceptions. &
\textit{``update: handle open ai not responding or failing in general with a default message''}
\\
\midrule

Performance Optimizations &
43 (2\%) &
Messages indicate optimizing the app's performance, such as improving loading times, reducing memory usage, and enhancing the efficiency of API calls. &
\textit{``Batch API calls when there is more than 1000 UIDs''} 
\\
\midrule

Localization and Language Support &
27 (1\%) &
Messages indicate efforts to add or improve localization features, such as adding translations and adjusting language settings. &
\textit{``Added support of multiple languages of voice assistant''}
\\
\midrule

Testing and Quality Assurance &
16 (1\%) &
Messages indicate testing activities, such as unit tests, integration tests, and quality assurance measures. &
\textit{``:white\_check\_mark: model and no stream completions has pass test''}
\\
\bottomrule

\end{tabular}
\end{table*}

\begin{figure}[htbp]
\centerline{\includegraphics[width=0.5\textwidth]{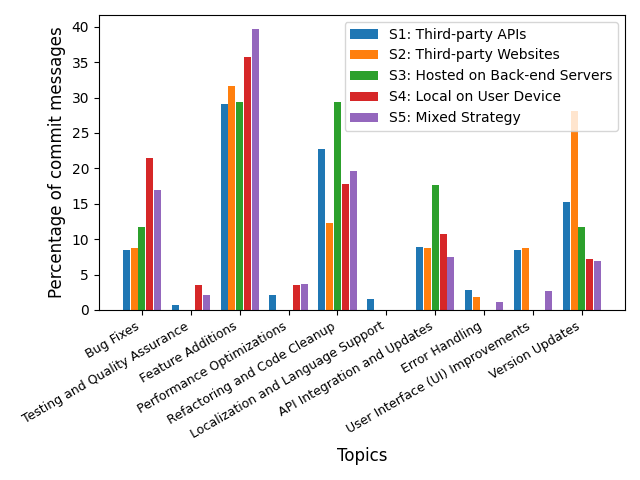}}
\vspace{-5mm}
\caption{The distribution of commit message topics per strategy.}
\label{CommitTopicsbyStrategy}
\end{figure}

\textbf{We find that \textit{“Feature Additions”} and \textit{“Refactoring and Code Cleanup”} occurs in the top 3 most common topics across all integration strategies.}
Tab.~\ref{commits_by_strategy} displays the total number of commits across all apps, per integration strategy. The percentage breakdown of commit messages assigned to each topic per integration strategy is shown in Fig.~\ref{CommitTopicsbyStrategy}. 
\textit{``Feature Additions''} is not only the most common topic across all commit messages, but also the most common topic for every integration strategy. 
As well, \textit{``Refactoring and Code Cleanup''} is also in the top 3 most common topics across all integration strategies. 
\textit{Localization and Language Support}, the second least common topic, is only present for apps that use third-party APIs as their integration strategy. 

\begin{table}[h]
\caption{The total number of commits per integration strategy.}
\label{commits_by_strategy}
\centering
\begin{tabular}{lc}
\toprule
\textbf{Integration Strategy}&{\textbf{Number of Commits}} \\
\midrule
IS1: Using third-party API calls & 1,712\\

\midrule

IS5: Mixing multiple integration strategies & 189 \\

\midrule

IS2: Using third-party website & 57 \\

\midrule

IS4: Hosting LLMs on user devices & 28 \\

\midrule

IS3: Hosting LLMs on back-end servers & 17 \\

\bottomrule
\end{tabular}
\vspace{-1.5em}
\end{table}

\section{Threats to Validity}
\label{threats}
 Regarding \emph{internal validity}, our study required manual effort in classifying apps as LLM-enabled and identifying LLM-related code, which introduces potential subjectivity. To mitigate this, multiple authors independently reviewed and crosschecked classifications, following predefined criteria to ensure consistency. 
 Additionally, our study's reproducibility is constrained by the use of an evolving LLM (GPT-4o-mini) for categorizing app functionality and commit message topics. Variability in LLM outputs over time and potential API changes could affect exact replication. To enhance \emph{reliability}, all LLM-based classifications were manually verified by the authors. About \emph{external validity}, our study  considered a limited set of LLMs, which may restrict the generalizability of our findings. The results may not fully extend to LLMs with different architectures, training data, or application domains. Future work could expand the analysis to a broader range of models to improve external validity.

\section{Conclusion}
\label{conclusion}
\looseness=-1
In this study, we present a novel dataset of 149 LLM-enabled Android apps and report an empirical investigation into the characteristics of this dataset, the strategies that developers use to integrate LLMs into their apps, and the reasons why developers may update LLM-related code as well as its implications on app releases. We identify five integration strategies used: using a third-party API, calling LLMs from a third-party website, hosting LLMs on a back-end server, hosting LLMs on the user end-device, and a hybrid strategy. Among them, LLM integration via API is the most common in our dataset, while storing LLMs locally on-device is the least common. 
We also investigate reasons why developers may update LLM-related code by using OpenAI's GPT-4o-mini to identify the top ten most common topics among all commit messages used for changes to LLM-related code. 

\section{Acknowledgments}
\label{Acknowledgments}
This work is partially supported by the Natural Sciences and Engineering Research Council of
Canada (NSERC), RGPIN-2021-03538 and RGPIN-2021-03969.



\bibliographystyle{IEEEtran}
\bibliography{references}

\end{document}

%% file: setup.tex
\usepackage{ragged2e}
\usepackage{multirow}
\usepackage{multicol}
\usepackage{amsmath}
\usepackage{mathtools}
\usepackage[most]{tcolorbox}
\usepackage{tablefootnote}
\usepackage{booktabs}
\usepackage{array}
\usepackage{colortbl}
\usepackage{color}
\usepackage{xcolor}
\usepackage{listings}
\usepackage{hyperref}
\usepackage{fontawesome5}
\usepackage{makecell}

\newcolumntype{L}[1]{>{\raggedright\let\newline\\\arraybackslash\hspace{0pt}}m{#1}}
\newcolumntype{C}[1]{>{\centering\let\newline\\\arraybackslash\hspace{0pt}}m{#1}}
\newcolumntype{R}[1]{>{\raggedleft\let\newline\\\arraybackslash\hspace{0pt}}m{#1}}

\newtcbox{\catbox}[1][red]{on line,
    colback=#1, colframe=#1, boxsep=0pt, boxrule=0pt, size=small, arc=0.7mm, left=-1pt, right=-1pt, top=-1pt, bottom=-0.5pt}

\newtcbtheorem{Summary}{\bfseries Summary of RQ}{enhanced,
	coltitle=black,
	top=0.1in,
	attach boxed title to top left=
	{xshift=1.5em,yshift=-\tcboxedtitleheight/2},
	boxed title style={size=small,colback=lightgray}
}{summary}

\newtcbtheorem{Summary_data_charac}{Summary}{enhanced,
	coltitle=black,
	top=0.1in,
	attach boxed title to top left=
	{xshift=1.5em,yshift=-\tcboxedtitleheight/2},
	boxed title style={size=small,colback=gray}
}{Summary_data_charac}

\newtcolorbox[auto counter]{summary}[1][]{title={\bfseries Summary},enhanced,
	coltitle=black,
	top=0.17in,
	attach boxed title to top left=
	{xshift=1.5em,yshift=-\tcboxedtitleheight/2},
	boxed title style={size=small,colback=lightgray},#1}

\newtcolorbox[auto counter]{summary_RQ1}[1][]{title={\bfseries Summary of RQ1},enhanced,
	coltitle=black,
	top=0.17in,
	attach boxed title to top left=
	{xshift=1.5em,yshift=-\tcboxedtitleheight/2},
	boxed title style={size=small,colback=lightgray},#1}

\newtcolorbox[auto counter]{summary_RQ2}[1][]{title={\bfseries Summary of RQ2},enhanced,
	coltitle=black,
	top=0.17in,
	attach boxed title to top left=
	{xshift=1.5em,yshift=-\tcboxedtitleheight/2},
	boxed title style={size=small,colback=lightgray},#1}

\newtcolorbox[auto counter]{summary_RQ3}[1][]{title={\bfseries Summary of RQ3},enhanced,
	coltitle=black,
	top=0.17in,
	attach boxed title to top left=
	{xshift=1.5em,yshift=-\tcboxedtitleheight/2},
	boxed title style={size=small,colback=lightgray},#1}